# PMC Physics B

Research article

# Effect of biased noise fluctuations on the output radiation of coherent beat laser

## Sintayehu Tesfa

Address: Physics Department, Dilla University, PO Box 419, Dilla, Ethiopia

Email: Sintayehu Tesfa - sint_tesfa@yahoo.com

## Abstract

Effect of biased noise fluctuations on the degree of squeezing as well as the intensity of a radiation generated by a one-photon coherent beat laser is presented. It turns out that the radiation exhibits squeezing inside and outside the cavity under certain conditions. The degree of squeezing is enhanced by the biased noise input significantly in both regions. Despite the presence of the biased environment modes outside the cavity, the degree of squeezing outside the cavity can be greater than or equal to or even less than the cavity radiation depending on the initial preparation of the atomic superposition and amplitude of the external driving radiation. But the intensity of the radiation is found to be lesser outside the cavity regardless of these parameters.

## Introduction

In recent years, interaction of three-level atoms with radiation has attracted a great deal of interest in relation to the strong correlation induced particularly during the cascading transitions [1-13]. It is common knowledge by now that the atomic coherence in such a system is accountable for the squeezing of the emitted radiation. The atomic coherence can be induced in a three-level cascade scheme by coupling the upper energy level $|a\rangle$ and lower energy level $|c\rangle$, between which a direct transition is dipole forbidden, with external radiation [1-6] or by preparing the atom initially in coherent superposition of these two levels [7-12] or using these mechanisms at the same time [13]. In addition to these options, Xu and Hu [14] have considered the two-step cascade coherent excitation. For the sake of convenience, the amplification of light when spontaneously emitted photons in the cascade transitions are correlated by the atomic coherence resulting from the initial preparation of the superposition and external driving mechanism can be taken as coherent beat laser (CBL). The initially prepared atoms are assumed to be injected into the cavity at constant rate and removed after they spontaneously decay to energy levels that are not involved in the lasing process. It is a well known fact that in three-level cascading process, basically, two photons are generated. If the two photons have identical frequency, the three-level atom is referred to as a degenerate three-level atom.

In actual experimental setting, the cavity is unavoidably coupled to the fluctuations in the external environment modes. As a result, the quantum properties in such a system is limited by the leakage through the mirror and inevitable amplification of the quantum fluctuations in the cavity. In other words, the squeezing of the cavity radiation in particular is degraded since the vacuum field has fundamentally no definite phase. In this accord, various authors have studied similar scheme coupled to vacuum reservoir when the atomic coherence is induced by external driving radiation and when the atoms are initially prepared to be in the upper energy level [1], lower energy level [2] and arbitrary coherent superposition of the two levels. It has been argued that the three-level laser in these cases resemble the corresponding parametric oscillator in the strong driving limit. However, replacing the ordinary vacuum reservoir by squeezed vacuum can enhance the squeezing of the cavity radiation [15,16]. Based on this, the idea of coupling the cavity radiation of phase-sensitive amplifier to biased noise fluctuations of broadband environment modes has been explained recently [7]. Since the squeezing in the phase-sensitive laser corresponds to unequal gain and unequal noise in the quadrature phases, coupling the cavity to biased noise fluctuations is expected to lead to enhancement of the degree of squeezing as long as the environment modes are biased in the right quadrature. In the present day state of the art technology, biased noise fluctuations can be generated, for instance, by optical feedback loop [17]. In light of this, the effects of the biased noise fluctuations (squashed light and twin beams) on the radiative properties of a three-level cascade atom have been discussed by Wiseman and co-worker [18-20] earlier.

Though previous works deal mainly with the degree of squeezing and statistical properties of the cavity radiation, it is the output radiation that is accessible to the experimenter and can readily be utilized. In this respect, it appears that there is renewed interest in comparing the squeezing and intensity of the radiation inside the cavity with the outside radiation using the input-output relation introduced by Gardiner and Collett [21]. It is found that the squeezing of the output radiation for the degenerate parametric oscillator coupled to broadband squeezed vacuum reservoir is less than the cavity radiation contrary to earlier expectations [22]. In addition, it has been shown recently that the output radiation can have a better degree of squeezing than the cavity radiation for degenerate correlated emission laser coupled to broadband squeezed vacuum reservoir (Tesfa S, unpublished data). Despite these efforts, experimental realization of the squeezed state indicates that the band of the squeezed light is typically in the order of the atomic line width [23] which automatically defies the broadband approximation. It, hence, seems imperative studying the squeezing of the radiation outside when the cavity is coupled to experimentally realizable broadband environment modes. That is why the output of the degenerate CBL whose cavity is coupled to broadband biased noise fluctuations is considered. Moreover, this work is confined to the regime of lasing without population inversion due to the limitation imposed by the uncer-tainty condition. The quadrature variances and mean photon number of the output radiation are calculated in the linear and adiabatic approximation schemes in the good cavity limit. The situation in which the atoms are initially prepared with equal probability to be in the upper and lower energy levels is taken as a particular case. An equal emphasis is given to the comparison of the squeezing and intensity of the output and cavity radiations. For clarity the schematic representation of the quantum system under consideration is given in Fig 1.

## Equation of evolution

Interaction of externally pumped cascade three-level atom with the cavity radiation can be described in the rotating-wave approximation and interaction picture by the Hamiltonian of the form

$$\hat{H}_{AR} = ig[\hat{a}(|a\rangle\langle b| + |b\rangle\langle c|) + (|b\rangle\langle a| + |c\rangle\langle b|)\hat{a}^{\dagger}] + i\frac{\Omega}{2}[|c\rangle\langle a| - |a\rangle\langle c|], \tag{1}$$

where $g$ is the coupling constant taken to be the same for both transitions, $\hat{a}$ is the annihilation operator for the cavity mode and $\Omega$ is a real-positive constant proportional to the amplitude of the driving radiation. The initial state of a three-level atom is taken to be $|\Phi_A(0)\rangle = C_a(0)|a\rangle + C_c(0)|c\rangle$, where $C_a(0)$ and $C_c(0)$ are the probability amplitudes for the atom to be in the upper and lower energy levels. This consideration corresponds to the case when the three-level atom is initially prepared to be in arbitrary coherent superposition of the upper and lower energy levels, in which the initial density operator for the atom would be

$$\hat{\rho}_A(0) = \rho_{aa}^{(0)}|a\rangle\langle a| + \rho_{ac}^{(0)}|a\rangle\langle c| + \rho_{ca}^{(0)}|c\rangle\langle a| + \rho_{cc}^{(0)}|c\rangle\langle c|. \tag{2}$$

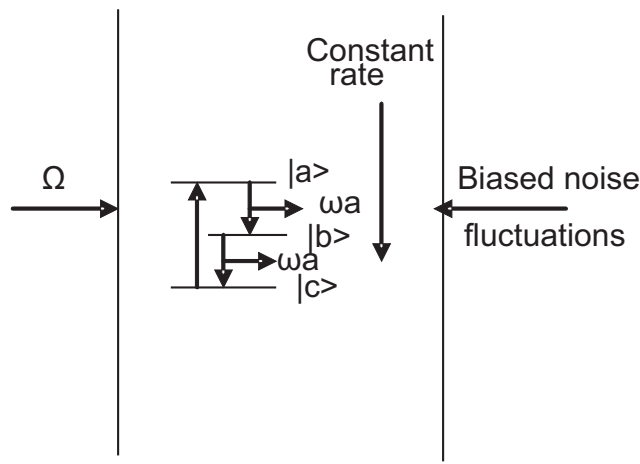

**Figure 1**
**Schematic representation of a coherently pumped degenerate three-level atom in a cascade configuration**. The transitions from $|a\rangle \rightarrow |b\rangle$ and from $|b\rangle \rightarrow |c\rangle$ at frequency $\omega_a$ are taken to be resonant with the cavity, whereas the transition from $|a\rangle \rightarrow |c\rangle$ is dipole forbidden. However, the transition from $|c\rangle \rightarrow |a\rangle$ is induced by pumping the atoms externally with a resonant radiation of frequency $2\omega_a$. Moreover, biased noise fluctuations enter the cavity via one of the coupler mirrors.

It is not difficult to notice that $\rho_{aa}^{(0)}$ and $\rho_{cc}^{(0)}$ are the probabilities for the atoms to be initially in the upper and lower energy levels, whereas $\rho_{ac}^{(0)}$ represents the initial atomic coherence.

It is assumed that the atoms are injected into the cavity at a constant rate $r_a$ and removed after sometime, which is long enough for the atoms to decay spontaneously to levels that do not contribute to the lasing process. The atomic spontaneous decay rate $\gamma$ is taken to be the same for the involved levels. Applying the linear and adiabatic approximation schemes in the good cavity limit [8], the time evolution of the density operator for the cavity mode of degenerate CBL coupled to broadband biased noise fluctuations via a single port-mirror is found using the standard method [24] along with the recently discussed idea [18,25] to be

$$
\begin{aligned}
\frac{d\hat{\rho}(t)}{dt} &= \frac{1}{2}\left( \frac{AC}{B} + \kappa N \right)[2\hat{a}^{\dagger}\hat{\rho}\hat{a} - \hat{a}\hat{a}^{\dagger}\hat{\rho} - \hat{\rho}\hat{a}\hat{a}^{\dagger}] \\
&+ \frac{1}{2}\left( \frac{AD}{B} + \kappa(N+1) \right)[2\hat{a}\hat{\rho}\hat{a}^{\dagger} - \hat{a}^{\dagger}\hat{a}\hat{\rho} - \hat{\rho}\hat{a}^{\dagger}\hat{a}] \\
&+ \frac{1}{2}\left( \frac{AE}{B} - \kappa M \right)[\hat{a}^{\dagger}\hat{\rho}\hat{a}^{\dagger} - \hat{a}^{2}\hat{\rho} - \hat{\rho}\hat{a}^{\dagger^{2}} + \hat{a}\hat{\rho}\hat{a}] \\
&+ \frac{1}{2}\left( \frac{AF}{B} - \kappa M \right)[\hat{a}^{\dagger}\hat{\rho}\hat{a}^{\dagger} - \hat{a}^{\dagger^{2}}\hat{\rho} - \hat{\rho}\hat{a}^{2} + \hat{a}\hat{\rho}\hat{a}],
\end{aligned}
\tag{3}
$$

where $\kappa$ is the cavity damping constant, $A = \frac{2r_a g^2}{\gamma^2}$ is the linear gain coefficient,

$$
B = \left( 1 + \frac{\Omega^2}{\gamma^2} \right)\left( 1 + \frac{\Omega^2}{4\gamma^2} \right),
\tag{4}
$$

$$
C = \rho_{aa}^{(0)}\left( 1 + \frac{\Omega^2}{4\gamma^2} \right) - \rho_{ac}^{(0)}\frac{3\Omega}{2\gamma} + \rho_{cc}^{(0)}\frac{3\Omega^2}{4\gamma^2},
\tag{5}
$$

$$
D = \rho_{aa}^{(0)}\frac{3\Omega^2}{4\gamma^2} + \rho_{ac}^{(0)}\frac{3\Omega}{2\gamma} + \rho_{cc}^{(0)}\left( 1 + \frac{\Omega^2}{4\gamma^2} \right),
\tag{6}
$$

$$
E = -\rho_{aa}^{(0)}\frac{\Omega}{2\gamma}\left( 1 - \frac{\Omega^2}{2\gamma^2} \right) - \rho_{ac}^{(0)}\left( 1 - \frac{\Omega^2}{2\gamma^2} \right) + \rho_{cc}^{(0)}\frac{\Omega}{\gamma}\left( 1 + \frac{\Omega^2}{4\gamma^2} \right),
\tag{7}
$$

$$
F = -\rho_{aa}^{(0)}\frac{\Omega}{\gamma}\left( 1 + \frac{\Omega^2}{4\gamma^2} \right) - \rho_{ac}^{(0)}\left( 1 - \frac{\Omega^2}{2\gamma^2} \right) + \rho_{cc}^{(0)}\frac{\Omega}{2\gamma}\left( 1 - \frac{\Omega^2}{2\gamma^2} \right).
\tag{8}
$$

It is worth noting that $N$ is the mean photon number of the environment mode capable of coupling with the cavity radiation. As a result, $N$ can be interpreted as the measure of the intensity of the biased noise fluctuations whereas $M = \sqrt{N(N+1)}$ as the degree of the correlation among the biased modes. Moreover, from the form of the master equation, one can realize that $C$ corresponds to the gain but $D$ to the lose of the cavity mode. On the other hand, $E$ and $F$ are related to the correlation of the generated radiation that indicates the existence of quantum features.

Making use of the master equation (3), the time evolution of the cavity mode in c-number variables associated with the normal ordering can be put following the procedure in as

$$\frac{d}{dt}\alpha(t) = -\frac{\mu}{2}\alpha(t) + \beta\alpha^*(t) + f(t), \qquad (9)$$

where $\alpha(t)$ is the c-number variable related to the cavity mode $\hat{a}(t)$ in the normal ordering, $\mu = \frac{A}{B}(D - C) + \kappa$, $\beta = \frac{A}{B}(E - F) + \kappa$ and $f(t)$ is the corresponding stochastic noise force with the correlation properties:

$$\langle f(t)\rangle = 0, \qquad (10)$$

$$\langle f(t')f(t)\rangle = -\left(\frac{AF}{B} - \kappa M\right)\delta(t - t'), \qquad (11)$$

$$\langle f(t)f^*(t')\rangle = \left(\frac{AC}{B} + \kappa N\right)\delta(t - t'). \qquad (12)$$

Upon introducing,

$$\alpha_\pm(t) = \alpha^*(t) \pm \alpha(t), \qquad (13)$$

one can easily see that

$$\frac{d}{dt}\alpha_\pm(t) = -\frac{\lambda_\mp}{2}\alpha_\pm(t) + f^*(t) \pm f(t), \qquad (14)$$

where

$$\lambda_\pm = \mu \mp 2\beta. \qquad (15)$$

Therefore, formal integration of Eq. (14) leads to

$$\alpha(t) = a_+(t)\alpha(0) + a_-(t)\alpha^*(0) + G_-(t) + G_+(t), \tag{16}$$

in which

$$a_\pm(t) = \frac{1}{2}\left( e^{-\frac{\lambda-t}{2}} \pm e^{-\frac{\lambda+t}{2}} \right), \tag{17}$$

$$G_\pm(t) = \frac{1}{2}\int_0^t e^{-\frac{\lambda\mp t}{2}}[f(t') \pm f^*(t')]dt'. \tag{18}$$

It is not difficult to notice that a well behaved solution of Eq. (14) exists at steady state provided that $\lambda > 0$. Hence $\mu = 2\beta$ is designated as a threshold condition. It perhaps worth mentioning that the squeezing as well as the statistical properties of the cavity radiation and output radiation are investigated with the aid of Eq. (16) when $\mu > 2\beta$.

### Quadrature variances

The squeezing of a single-mode output radiation can be described in terms of the quadrature operators defined by

$$\hat{a}_+^{out} = \hat{a}_{out}^\dagger + \hat{a}_{out} \tag{19}$$

and

$$\hat{a}_-^{out} = i(\hat{a}_{out}^\dagger - \hat{a}_{out}). \tag{20}$$

The variances of these operators can be put in the form

$$\Delta a_{\pm(out)}^2 = 1 \pm [\langle a_{out}^{\;2}(t)\rangle + \langle a_{out}^2(t)\rangle \pm 2\langle a_{out}(t)a_{out}(t)\rangle \\ + \langle \hat{a}_{out}^\dagger(t)\rangle^2 + \langle \hat{a}_{out}(t)\rangle^2 \pm 2\langle \hat{a}_{out}^\dagger(t)\rangle\langle \hat{a}_{out}(t)\rangle]. \tag{21}$$

It is straight forward to see that the operators in Eq. (21) are put in the normal order. Hence the corresponding expression in terms of the c-number variables associated with the normal ordering would be

$$\Delta a_{\pm(out)}^2 = 1 \pm [\langle \alpha_{out}^{*2}(t)\rangle + \langle \alpha_{out}^2(t)\rangle \pm 2\langle \alpha_{out}^*(t)\alpha_{out}(t)\rangle \\ + \langle \alpha_{out}^*(t)\rangle^2 + \langle \alpha_{out}(t)\rangle^2 \pm 2\langle \alpha_{out}^*(t)\rangle\langle \alpha_{out}(t)\rangle]. \tag{22}$$

Quite generally, the output radiation in the normal ordering can be represented in terms of the cavity mode variables as [21]

$$\alpha_{out}(t) = \sqrt{\kappa}\,\alpha(t) - \frac{1}{\sqrt{\kappa}}\,F_R(t),$$  (23)

where $F_R(t)$ is the noise force associated with the environment modes and satisfies for broadband biased noise fluctuations the correlations:

$$\langle F_R(t) \rangle = 0,$$  (24)

$$\langle F_R^*(t) F_R(t') \rangle = \kappa N \delta(t - t'),$$  (25)

$$\langle F_R(t)\, F_R(t') \rangle = \kappa M\ \delta(t - t').$$  (26)

In view of Eq. (23), when the cavity mode is taken to be initially in the vacuum state, Eq. (22) reduces to

$$\Delta a_{\pm(out)}^2 = 1 + 2\langle \alpha_{out}^*(t)\alpha_{out}(t) \rangle \pm [\langle \alpha_{out}^{*2}(t) \rangle + \langle \alpha_{out}^2(t) \rangle].$$  (27)

Moreover, one can write using Eq. (23) that

$$\langle \alpha_{out}^*(t)\alpha_{out}(t) \rangle = \kappa \langle \alpha^*(t)\alpha(t) \rangle + \frac{1}{\kappa}\langle F_R^*(t)F_R(t) \rangle - \langle \alpha^*(t)F_R(t) \rangle - \langle F_R^*(t)\alpha(t) \rangle.$$  (28)

It is possible to realize that the noise force related to the stochastic process ($f(t)$) represents the contribution of the vacuum fluctuations of the cavity as well as the environment modes and hence can be put in the form

$$f(t) = F_C(t) + F_R(t),$$  (29)

where $F_C(t)$ is the noise force corresponding to the system in the cavity in the absence of the biased input. On the basis of the fact that the noise force of the environment $F_R(t)$ does not correlate with $F_C(t)$ and the system variables at the earlier times along with the fact that the contribution of the biased noise fluctuations should be taken from initial time $t = 0$ to $t = \infty$, one can verify at steady state that

$$\langle F_R(t)\alpha^*(t) \rangle_{ss} + \langle F_R^*(t)\alpha(t) \rangle_{ss} = \kappa N.$$  (30)

Thus on account of Eqs. (25), (28) and (30), one finds

$$\langle \alpha_{out}^*(t)\alpha_{out}(t) \rangle_{ss} = \kappa \langle \alpha^*(t)\alpha(t) \rangle_{ss} + N\langle 1 - \kappa \rangle.$$  (31)

This is the mean photon number of the output radiation in which the first term is the contribution of the cavity radiation that escapes through the mirror, the second term is the mean photon number of the environment modes and the third term is the measure of the intensity of the biased noise fluctuations entering the cavity. It can also be readily obtained that

$$\langle \alpha_{out}^2(t) \rangle_{ss} = \kappa \langle \alpha^2(t) \rangle_{ss} + M(1 - \kappa). \tag{32}$$

Now making use of Eqs. (27), (31) and (32) results at steady state

$$\Delta a_{\pm(out)}^2 = (1 - \kappa)(1 + 2N \pm 2M) + \kappa \Delta a_{\pm}^2, \tag{33}$$

where $\Delta a_{\pm}^2$ are the variances of the cavity quadrature operators that can be expressed in terms of the corresponding c-number variables associated with the normal ordering as

$$\Delta a_{\pm}^2 = 1 \pm \langle \alpha_{\pm}^2(t) \rangle. \tag{34}$$

On the other hand, taking Eqs. (11), (12), (13) and (14) into consideration leads to

$$\langle \alpha_{\pm}^2(t) \rangle = -\frac{2}{\lambda_{\mp}} \left[ \frac{A}{B}(F \mp C) + \kappa(M \mp N) \right] (1 - e^{-\lambda_{\mp} t}). \tag{35}$$

In order to express Eq. (35) in a more convenient form, it appears customary to introduce a parameter

$$\rho_{aa}^{(0)} = \frac{1 - \eta}{2}, \tag{36}$$

with $-1 \leq \eta \leq 1$. It is straight forward to verify when the three-level atoms are initially prepared to be in arbitrary atomic coherent superposition that $\rho_{cc}^{(0)} = \frac{1+\eta}{2}$ and $\rho_{ac}^{(0)} = \frac{\sqrt{1-\eta^2}}{2}$. Therefore, on the basis of Eqs. (4), (5), (6), (7), (8), (15) and (36), Eq. (35) finally takes at steady state the form

$$\langle \alpha_{\pm}^2(t) \rangle_{ss} = \frac{A \left[ \frac{\Omega}{2\gamma} \left( 1 - 3\eta + \frac{\Omega^2}{\gamma^2} \right) + \sqrt{1-\eta^2} \left( 1 - \frac{\Omega^2}{2\gamma^2} \right) \right]}{\kappa H_{\pm}}$$

$$\pm \frac{A \left[ 1 - \eta + \frac{\Omega^2}{2\gamma^2}(2+\eta) - \sqrt{1-\eta^2} \frac{3\Omega}{2\gamma} \right]}{\kappa H_{\pm}} + \frac{2B(M \pm N)}{H_{\pm}}, \tag{37}$$

where

$$H_{\pm} = \left(1 + \frac{\Omega^2}{\gamma^2}\right)\left(1 + \frac{\Omega^2}{4\gamma^2}\right) + \frac{A}{\kappa}\left[\left(1 - \frac{\Omega^2}{2\gamma^2}\right)\eta\right.$$
$$\left. + \sqrt{1 - \eta^2}\,\frac{3\Omega}{2\gamma} \mp \frac{\Omega}{2\gamma}\left(1 + \frac{\Omega^2}{\gamma^2}\right)\right].$$

(38)

Hence the variances of the quadrature operators for the output radiation at steady state turn out to be

$$\Delta a^2_{\pm(out)} = \frac{\kappa H_{\pm} + A\left[1 - \eta + \frac{\Omega^2}{2\gamma^2}(2 + \eta) - \sqrt{1 - \eta^2}\,\frac{3\Omega}{2\gamma}\right]}{H_{\pm}}$$

$$\pm \frac{A\left[\frac{\Omega}{2\gamma}\left(1 - 3\eta + \frac{\Omega^2}{\gamma^2}\right) + \sqrt{1 - \eta^2}\left(1 - \frac{\Omega^2}{2\gamma^2}\right)\right]}{H_{\pm}}$$

$$+ \frac{2\kappa B(N \pm M)}{H_{\pm}} + (1 - \kappa)(1 + 2N \pm 2M).$$

(39)

This reduces for $A = 0$ to $\Delta a^2_{\pm(out)} = 1 + 2N \pm 2M$, since there is only biased noise fluctuations in the cavity in this case.

It has been shown elsewhere that the radiation generated by similar schemes exhibits squeezing only for certain values of $\Omega/\gamma$ and $\eta$, where the degree of squeezing generally found to increase with the linear gain coefficient. In order to analyze the dependence of the degree of squeezing of the output radiation on the intensity of the environment modes, amplitude of the driving radiation and injected atomic coherence more closely, it is necessary to consider various cases of interest separately. To this effect, for $\Omega = 0$

$$\Delta a^2_{\pm(out)} = \frac{\kappa^2(1 + 2N \pm 2M) + \kappa A(1 \pm \sqrt{1 - \eta^2})}{A\eta + \kappa} + (1 - \kappa)(1 + 2N \pm 2M). \qquad (40)$$

Despite the fact that a higher degree of squeezing is achievable for larger values of $A$, the linear gain coefficient is limited to smaller values so that the dependence of the degree of squeezing on the parameters under consideration is evident from the figures.

As clearly indicated in Fig. 2, the degree of squeezing of the output radiation increases with the intensity of the biased noise fluctuations in which a higher degree of squeezing is found for relatively smaller values of $\eta$. Even though the linear gain coefficient is limited to smaller values

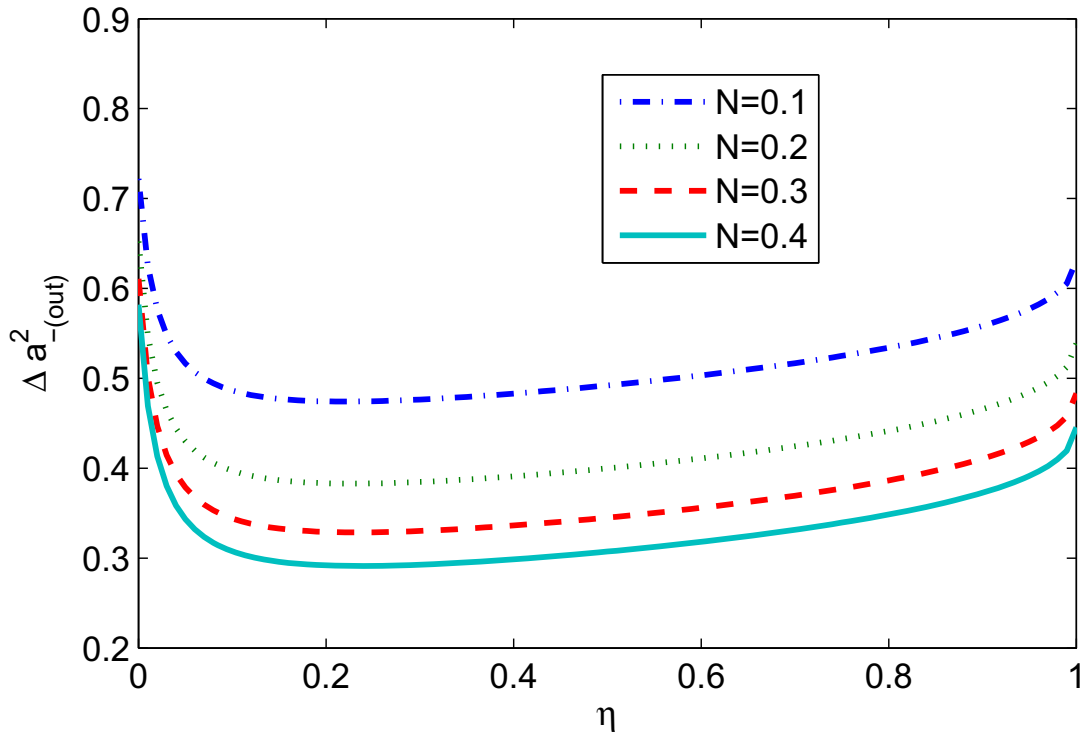

**Figure 2**

**Plots of the minus quadrature variance of the output radiation ($\Delta a^2_{-(out)}$) at steady state for $\kappa$ = 0.2, $\Omega$ = 0, $A$ = 10 and different values of $N$.**

for convenience, squeezing of 71% occurs for $N$ = 0.4 and $A$ = 10 at $\eta$ = 0.25 whereas for $N$ = 0.1 squeezing of 53% occurs at $\eta$ = 0.22. Hence it is observed that the biased noise fluctuations enhance the degree of squeezing of the output radiation substantially. It is believed that the correlated emission initiated by the initially prepared atomic coherent superposition is accountable for the reduction of the fluctuations of the noise in one of the quadrature components below the classical limit in addition to the biased input, since three-level laser falls under phase-sensitive amplifier [7,26].

Moreover, when the atoms are initially prepared with a maximum possible injected atomic coherence ($\eta$ = 0), one gets

$$\Delta a^2_{\pm(out)} = \frac{\kappa H_\pm \pm A\left[\frac{\Omega}{2\gamma}\left(\frac{\Omega^2}{2\gamma^2} - 2 - \frac{\Omega}{\gamma}\right) + 1\right]}{H''_\pm} + \frac{A\left[1 + \frac{\Omega^2}{\gamma^2} - \frac{3\Omega}{2\gamma}\right]}{H''_\pm}$$
$$+ \frac{2\kappa B(N \pm M)}{H''_\pm} + (1 - \kappa)(1 + 2N \pm 2M),$$

(41)

in which

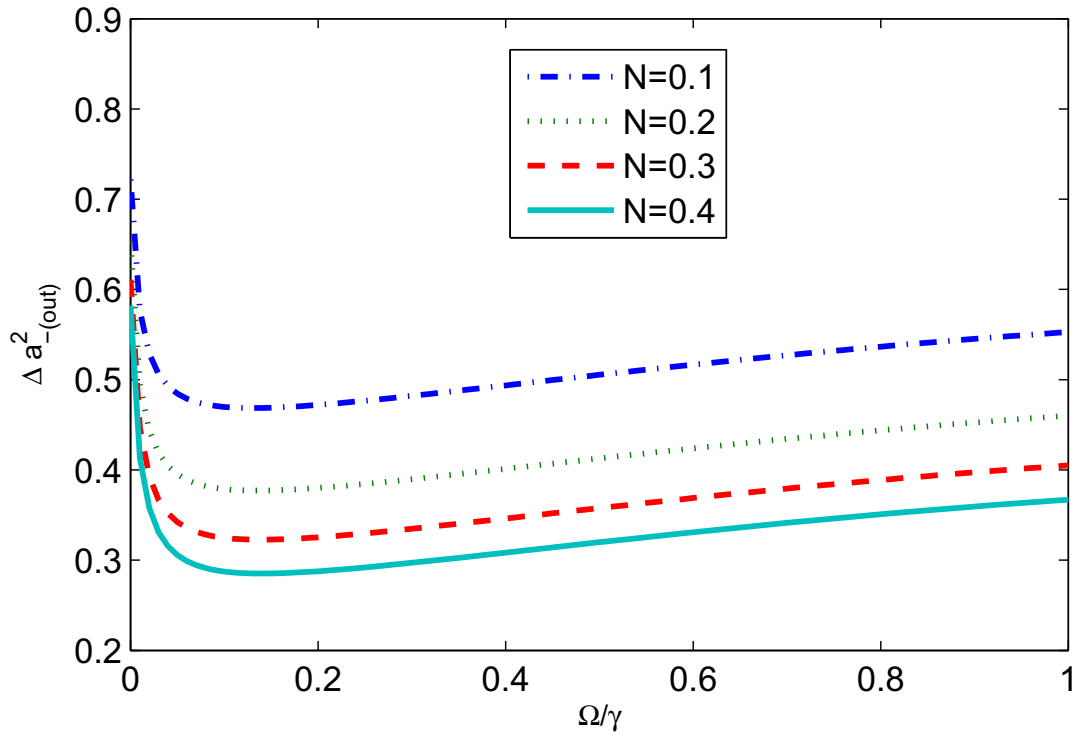

**Figure 3**

**Plots of the minus quadrature variance of the output radiation ($\Delta a^2_{-(out)}$) at steady state for $\kappa$ = 0.2, $\eta$ = 0, $A$ = 10 and different values of $N$.**

$$H''_{\pm} = \left(1 + \frac{\Omega^2}{\gamma^2}\right)\left(1 + \frac{\Omega^2}{4\gamma^2}\right) + \frac{A}{\kappa}\left[\frac{3\Omega}{2\gamma} \mp \frac{\Omega}{2\gamma}\left(1 + \frac{\Omega^2}{\gamma^2}\right)\right]. \qquad (42)$$

It is not difficult to see from Fig. 3 that the output radiation of the system under consideration, when the atoms are initially prepared with a maximum injected coherence, exhibits a substantial degree of squeezing for smaller values of $\Omega/\gamma$. It is found that a maximum squeezing of 71% occurs at $\Omega = 0.14\gamma$ for $A = 10$ and $N = 0.4$. The degree of squeezing decreases with the amplitude of the driving radiation for larger values of $\Omega/\gamma$, but it increases with the intensity of the biased noise fluctuations throughout. Further manipulation reveals that squeezing of higher degree than 71% can be realizable in the system under consideration when the atoms are initially prepared in a possible maximum coherent superposition and externally pumped with radiation of relatively smaller amplitude. It is expected that the possibility for generating highly squeezed light by altering various parameters will make this system reliable and attractive source of squeezed light.

In the following, the minus quadrature variance would be plotted for the cavity radiation and output radiation so that comparison between the squeezing inside and outside the cavity can be made. It can be shown for $A = 0$ that the degree of squeezing inside and outside the cavity is the same, since there is basically the same radiation in both regions. It is not difficult to see from Figs. 4 and 5 that the degree of squeezing outside the cavity can be greater than that inside the cavity for certain values of the initial preparation of the atomic coherent superposition. It is believed that this can be taken as one of the essential encouragements to utilize the generated squeezed radiation outside the cavity. It can also be deduced that when compared to the cavity radiation, the output radiation slowly varies with the initial preparation of the superposition and amplitude of the external radiation. Since the output radiation is the superposition of the radiation escaping through the coupler mirror and biased environment modes reflected from the same mirror, its degree of squeezing is undoubtedly influenced by the biased fluctuations more significantly. This must be the reason for getting nearly a uniform degree of squeezing for the output radiation close to the squeezing of the environment modes.

On the basis of the definition of the parameter $\eta$ (Eq. (36)), it can be realized for $\eta = 0$ that $\rho_{aa}^{(0)} = \rho_{cc}^{(0)} = \rho_{ac}^{(0)} = 1/2$, which corresponds to a maximum possible initial atomic coherence. But a case for which $\eta = 1$, $\rho_{aa}^{(0)} = \rho_{ac}^{(0)} = 0$ and $\rho_{cc}^{(0)} = 1$, is related to the absence of injected atomic coherence at the beginning. One can readily see from Eq. (40) that there is no squeezing when the atoms are initially prepared with a maximum or minimum atomic coherence, if they are not pumped externally ($\Omega = 0$) and there is no biased environment ($N = 0$). However, as shown in Figs. 2 and 4, the maximum squeezing occurs when the atoms are prepared with initial coherence close to the maximum possible value. It can also be observed from Figs. 3 and 4 that the external driving radiation initiates correlation between the states of the emitted radiation in the cascading process which leads to squeezing inside and outside the cavity for $\eta = 0$ even when $N = 0$. When there is no injected atomic coherence ($\eta = 1$) and $N = 0$, squeezing close to 50% is obtained for a very large amplitude of the external radiation ($\Omega^- \gamma$). This result agrees with the recent claim that three-level laser in which the atoms are initially prepared in the bottom level and externally pumped by strong radiation resembles parametric oscillator [2,14]. Moreover, comparison of the results shown in Figs. 2 and 3 reveals that a higher degree of squeezing can be obtained when the atoms are initially prepared with maximum atomic coherence and pumped externally with radiation of relatively smaller amplitude. Likewise, a maximum noise reduction when the atoms are injected into a resonant cavity with a maximum atomic coherence and pumped with external radiation for $\Lambda$ three-level laser has been shown [3]. Though the external radiation induces the atomic coherence accountable for the squeezing, it is observed that pumping the atoms with a stronger radiation than required destroys the squeezing. For parameters under consideration the degree of squeezing is found to increase with the intensity of the biased noise fluctuations, however as recent calculation shows the degree of squeezing in the nondegenerate case can decrease with the intensity of the biased fluctuations [7] due to the phenomenon of electron shelving [27,28].

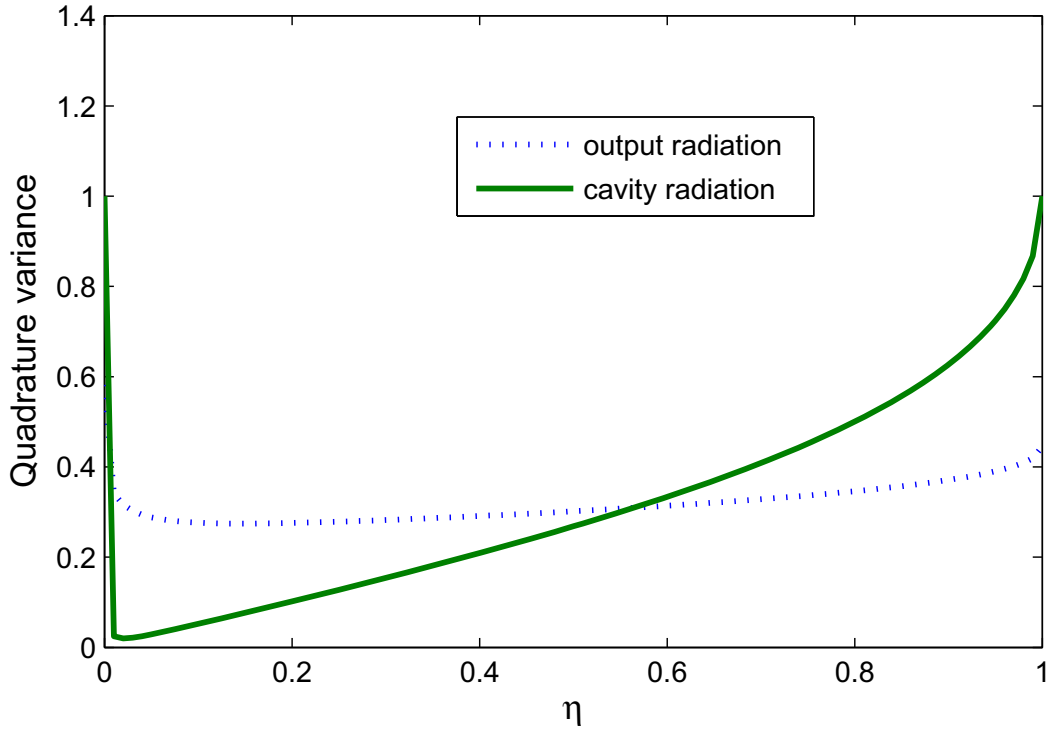

**Figure 4**
**Plots of the output and cavity minus quadrature variances at steady state for $\kappa$ = 0.2, $\Omega$ = 0, $A$ = 1000 and $N$ = 0.4.**

## Mean photon number

The mean photon number of the output radiation can be defined as

$$\bar{n}_{out} = \langle \alpha_{out}^{*}(t)\alpha_{out}(t)\rangle, \tag{43}$$

which can be expressed with the aid of Eqs. (13) and (31) at steady state in the form

$$\bar{n}_{out} = \kappa \left[ \frac{\langle \alpha_{+}^{2}(t)\rangle - \langle \alpha_{-}^{2}(t)\rangle}{4} \right] + N(1-\kappa). \tag{44}$$

Hence on account of Eq. (37) one obtains

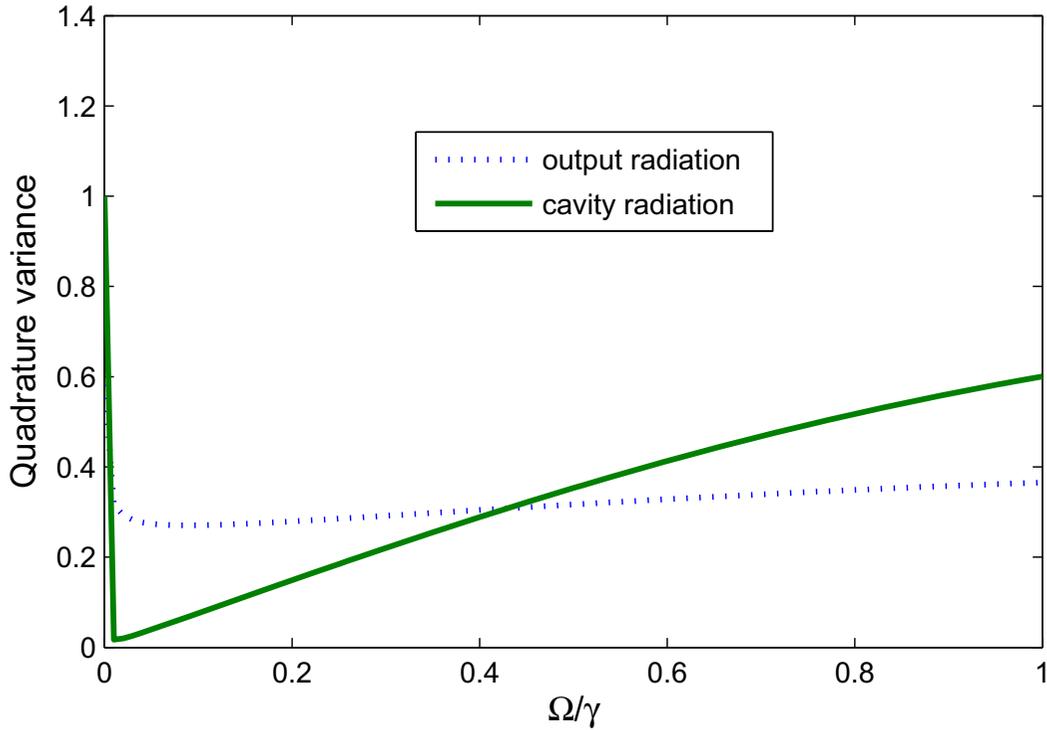

**Figure 5**
**Plots of the output and cavity minus quadrature variances at steady state for** $\kappa = 0.2$, $\eta = 0$, $A = 1000$ **and** $N = 0.4$.

$$
\begin{aligned}
\bar{n}_{out} = {} & \frac{A\sqrt{1-\eta^2}\left(\dfrac{\Omega^2}{2\gamma^2}-1-\dfrac{3\Omega}{2\gamma}\right)+2\kappa B(N-M)}{4H_-} \\[2ex]
& + \frac{A\left[\dfrac{\Omega}{2\gamma}(1-3\eta)+\dfrac{\Omega^3}{2\gamma^3}+\left(1-\eta+\dfrac{\Omega^2}{2\gamma^2}(2+\eta)\right)\right]}{4H_+} \\[2ex]
& - \frac{A\left[\dfrac{\Omega}{2\gamma}(1-3\eta)+\dfrac{\Omega^3}{2\gamma^3}+\left(1-\eta+\dfrac{\Omega^2}{2\gamma^2}(2+\eta)\right)\right]}{4H_-} \\[2ex]
& - \frac{A\sqrt{1-\eta^2}\left(\dfrac{\Omega^2}{2\gamma^2}-1+\dfrac{3\Omega}{2\gamma}\right)-2\kappa B(N+M)}{4H_+} + N(1-\kappa).
\end{aligned}
\tag{45}
$$

It can be inferred from Eq. (45) that radiation can be generated when initially all atoms are prepared to be in the lower energy level and if there is no external driving radiation contrary to previous reports, since the atoms can absorb the biased input to be excited to the upper energy levels from which emission of the required radiation is possible. It is not difficult to notice that, as a result of the external pumping, it is possible to produce a strong radiation from the laser even when the atoms are initially prepared to be in the lower energy level. This demonstrates that the mechanism of lasing without population inversion is readily evident in this scheme.

In order to study the dependence of the output mean photon number on the intensity of the biased noise fluctuations and linear gain coefficient, some cases of interest are considered. For instance, when $\Omega = 0$ Eq. (45) reduces to

$$\bar{n}_{out} = \frac{\kappa A(1-\eta)+2\kappa^2 N}{2(A\eta+\kappa)} + N(1-\kappa). \tag{46}$$

It is possible to see that the mean photon number would be zero if there is no driving radiation and all the atoms are initially prepared to be in the lower energy level when $N = 0$. One gets the strongest radiation when all atoms are initially prepared to be in the upper energy level as expected. In addition, for $\eta = 0$ Eq. (45) takes the form

$$\begin{aligned}
\bar{n}_{out} = &\frac{A\left[\dfrac{\Omega}{2\gamma}\left(1+\dfrac{\Omega^2}{\gamma^2}+\dfrac{\Omega}{\gamma}\right)-2-\left(\dfrac{3\Omega}{2\gamma}+\dfrac{\Omega^2}{\gamma^2}\right)\right]}{4H''_+} \\
&-\frac{A\left[\dfrac{\Omega}{2\gamma}\left(1+\dfrac{\Omega^2}{\gamma^2}+\dfrac{\Omega}{\gamma}\right)-\dfrac{3\Omega}{2\gamma}+\dfrac{\Omega^2}{\gamma^2}\right]-2\kappa B(N-M)}{4H''_-} \\
&+\frac{2\kappa B(N+M)}{4H''_+} + N(1-\kappa).
\end{aligned} \tag{47}$$

As clearly shown in Fig. 6, the intensity of the produced radiation decreases with the amplitude of the driving radiation if the atoms are initially prepared with a maximum possible atomic coherence for $\Omega < \gamma$. Though there are biased noise environment modes outside the cavity, as can readily be seen from Fig. 7, the mean photon number of the radiation in the cavity is much greater than the corresponding output radiation in many cases. This is actually related to the fact that in the good cavity limit ($\kappa$ small) much of the cavity radiation stays in the cavity. It is also clearly shown in Fig. 6 that the mean photon number increases with the intensity of the biased noise fluctuations. Even though it has been shown in previous work that the mean photon number increases with the linear gain coefficient, the values of $A$ are limited to smaller values so that the dependence of the mean photon number on the intensity of the biased noise fluctuations and amplitude of the driving radiation is evident from the figures. It is possible in principle to generate a strong radiation from the system under consideration by varying the rate at which the atoms are injected into the cavity, intensity of the biased environment modes, amplitude of the driving radiation and the way in which the atoms are initially prepared. Further manipulation reveals that there is a possibility for the intensity of the radiation to be equal inside and outside the cavity for some values of the parameters under consideration.

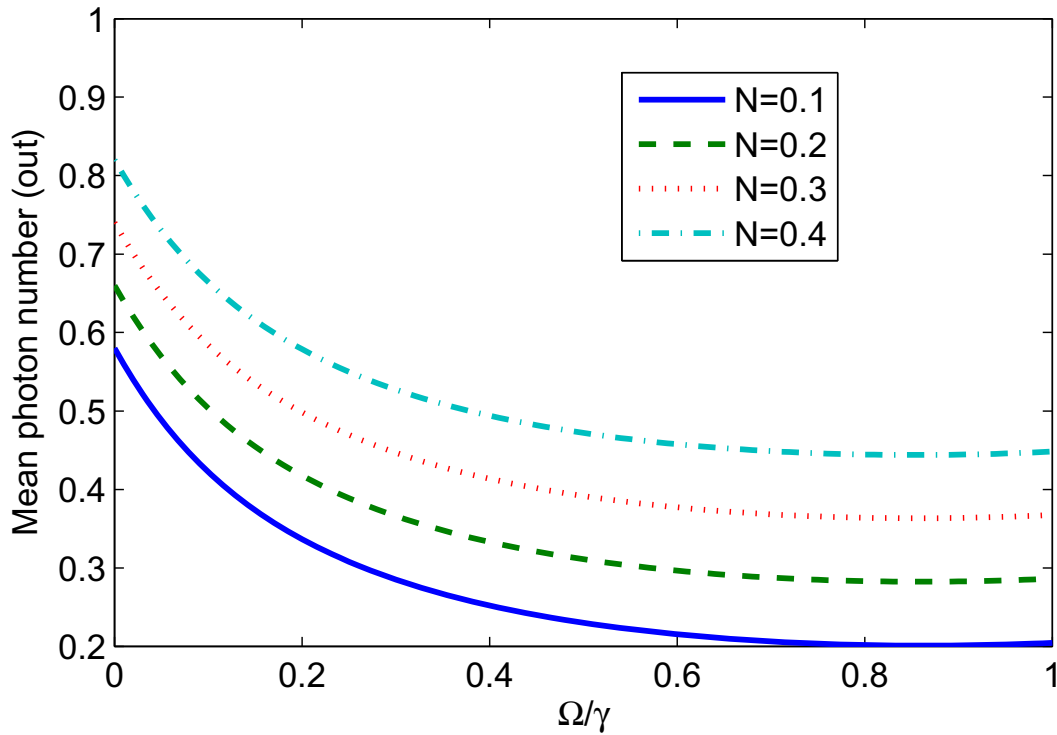

**Figure 6**
**Plots of the mean photon number of the output radiation at steady state for $\kappa$ = 0.2, $A$ = 1, $\eta$ = 0 and different values of $N$.**

## Conclusion

Detailed analysis of the degree of squeezing and intensity of the output radiation generated by a degenerate coherent beat laser coupled to broadband biased environment modes via one of the coupler mirrors is presented. It is found that the output as well as the cavity radiation exhibits squeezing under certain conditions pertaining to the initial preparation of the atomic coherent superposition, strength of the driving radiation and intensity of the biased fluctuations. Though the external driving radiation induces the atomic coherence accountable for the squeezing, pumping the atoms externally with a strong radiation results considerable reduction in the degree of squeezing and intensity of the radiation in both regions. Hence I cannot see the practical advantage of pumping the atoms with a strong radiation in this respect. That is why this study is confined mainly to weaker driving regime. Nonetheless intense radiation with a substantial degree of squeezing can be generated by the driving mechanism specially when the atoms are initially prepared with equal probability to be in the lower and upper energy levels, where there is no squeezing in the absence of the driving radiation and when $N$ = 0. It is also shown that it is possible to get squeezed radiation when the atoms are initially prepared to be in the lower energy level, where there is no radiation at all in the absence of driving process and $N$ = 0. It can be real-ized that driving mechanism can be considered as an option for generating squeezed radiation when it is technically difficult to prepare the atoms initially in arbitrary coherent superposition. Due to the additional atomic coherence induced by the driving process, it is reasonable to expect the enhancement of the degree of squeezing when the atoms initially prepared with arbitrary atomic coherent superposition are externally pumped; the fact that cannot be confirmed in this study for all cases.

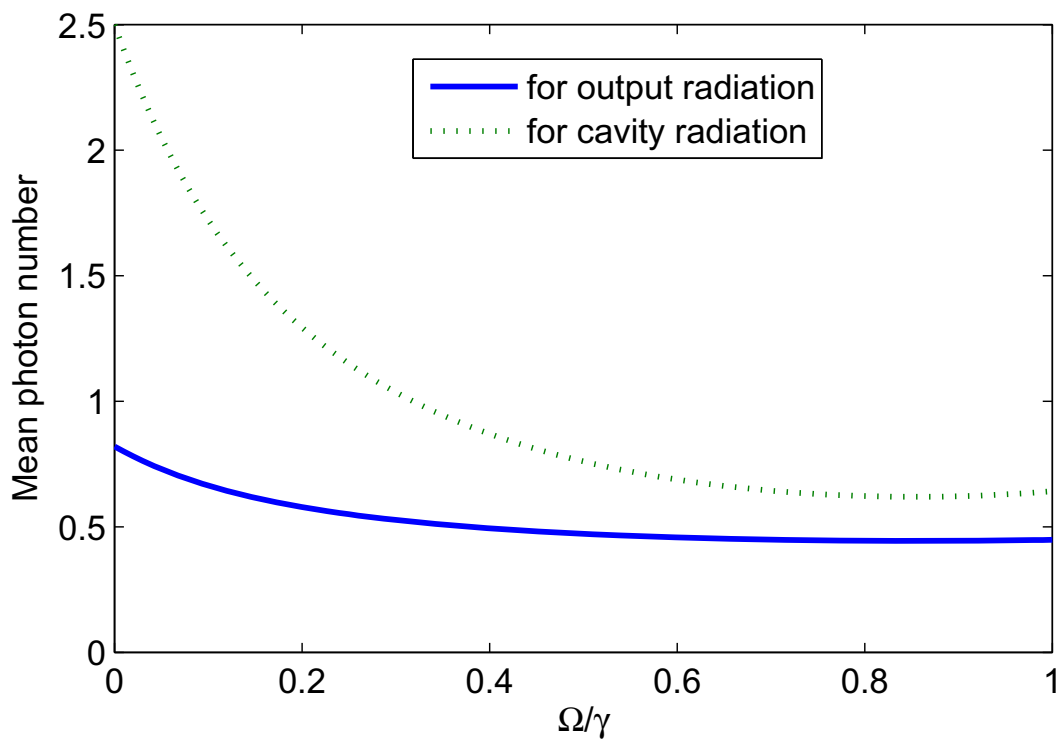

**Figure 7**
**Plots of the mean photon number of the cavity and output radiations at steady state for $\kappa$ = 0.2, $N$ = 0.4 and $A$ = 1.**

On the other hand, coupling the cavity to broadband biased environment modes is found to improve the degree of squeezing and intensity of the output as well as the cavity radiation. Contrary to this, the degree of squeezing in the nondegenerate three-level cascade laser does not always increase with the intensity of the biased noise fluctuations due to the phase competition in the correlation resulting from two different modes [7]. Comparison of the squeezing inside and outside the cavity reveals that the squeezing of the output radiation can be greater than or equal to or less than the cavity radiation depending on the parameters under consideration. Since the effect of the biased noise fluctuations would be prominent when there is small number of emitted photons, the degree of squeezing of the output radiation would be larger than the cavity radiation when relatively fewer atoms are initially prepared to be in the upper energy level. It is also found that though the mean photon number of the output radiation can be close to the cavity radiation, the intensity of the radiation in the cavity is much better in many instances. Hence despite the envisaged practical challenges to utilize the system under consideration, it is believed that the possibility of getting a strong squeezed light outside the cavity is not something that one brushes off without trying.

## References

1.   Ansari NA, Banacloche JG, Zubairy MS: *Phys Rev A* 1990, **41**:5179.

2.   Xiong H, Scully MO, Zubairy MS: *Phys Rev Lett* 2005, **94**:023601.

3.   Saaverda C, Retamal JC, Keitel CH: *Phys Rev A* 1997, **55**:3802.

4.   Martinez MAG, Herczfeld PR, Samuels C, Narducci LM, Keitel CH: *Phys Rev A* 1997, **55**:4483.

5.   Zhu Y: *Phys Rev A* 1997, **55**:4568.

6.   An S, Sargent M III: *Phys Rev A* 1989, **39**:1841.

7.   Tesfa S: *Phys Rev A* 2008, **77**:013815.

8.   Tesfa S: *Phys Rev A* 2006, **74**:043816.

9.   Tesfa S: *J Phys B: At Mol Opt Phys* 2006, **40**:2373.

10.  Ansari NA: *Phys Rev A* 1992, **46**:1560.

11.  Scully MO, Wodkiewicz K, Zubairy MS, Bergou J, Lu N, Meyer ter Van J: *Phys Rev lett* 1988, **60**:1832.

12.  Anwar J, Zubairy MS: *Phys Rev A* 1994, **49**:481.

13.  Ansari NA: *Phys Rev A* 1993, **48**:4686.

14.  Hu X, Xu Z: *J Phys B: At Mol Opt Phys* 2001, **34**:787.

15.  Banacloche JG: *Phys Rev Lett* 1987, **59**:543.

16.  Anwar J, Zubairy MS: *Phys Rev A* 1992, **45**:1804.

17.  Dariano GM, Sacchi M: *Mod Phys Lett B* 1997, **11**:1263.

18.  Wiseman HM: *Phys Rev Lett* 1998, **81**:3840.

19.  Wiseman HM: *J Opt B: Quantum Semiclassic Opt* 1999, **1**:459.

20.  Thomsen LK, Wiseman HM: *Phys Rev A* 2001, **64**:043805.

21.  Gardiner CW, Collett MJ: *Phys Rev A* 1985, **31**:3761.

22.  Tesfa S: *Nonl Opt Quant Opt* 2008, **38**:39.

23.  Georgiades NP, Plozik ES, Edamatsu K, Kimble HJ, Parkins AS: *Phys Rev Lett* 1995, **75**:3426.

24.  Louisell WH: *Quantum statistical properties of radiation* Wiley, Newyork; 1973.

25.  Tanas RJ: *Opt B: Quantum Semiclassic Opt* 2002, **4**:142.

26.  Caves CM: *Phys Rev D* 1982, **26**:1817.

27.  Buhner V, Tamm C: *Phys Rev A* 2000, **61**:061801(R).

28.  Evers J, Keitel CH: *Phys Rev A* 2002, **65**:033813.